\documentclass[sigconf]{acmart}

\usepackage{booktabs} 

\setcopyright{rightsretained}

\newcommand{\ignore}[1]{}
\usepackage{fancyhdr}
\usepackage[normalem]{ulem}
\usepackage[ruled,vlined,linesnumbered]{algorithm2e}
\usepackage[shortlabels]{enumitem} 
\usepackage{indentfirst}
\setlist[enumerate]{topsep=5pt,parsep=1pt}
\setlist[itemize]{topsep=2pt,parsep=1pt}
\let\oldnl\nl
\newcommand{\nonl}{\renewcommand{\nl}{\let\nl\oldnl}}

\usepackage[labelfont=bf]{caption}
\usepackage{subcaption}
\usepackage{multirow}
\usepackage{booktabs,tabularx}
\newcommand{\ra}[1]{\renewcommand{\arraystretch}{#1}}
\newcolumntype{L}[1]{>{\raggedright\arraybackslash}p{#1}}
\newcolumntype{C}[1]{>{\centering\arraybackslash}p{#1}}
\newcolumntype{R}[1]{>{\raggedleft\arraybackslash}p{#1}}

\acmISBN{}

\acmConference[]{xxx}{xxx}{xxx}
\acmYear{2018}
\copyrightyear{2018}

\acmArticle{4}
\acmPrice{15.00}

\begin{document}
\title{Gemini: Reducing DRAM Cache Hit Latency by Hybrid Mappings}

\author{Ye  Chi}
\affiliation{%
  \institution{Huazhong University of Science and Technology}
  \city{Wuhan}
  \state{China}
}

\begin{abstract}

	Die-stacked DRAM caches are increasingly advocated to bridge the performance gap between on-chip Cache and main memory.  It is essential to improve DRAM cache hit rate and lower cache hit latency simultaneously. Prior DRAM cache designs fall into two categories according to the data mapping polices: set-associative and direct-mapped, achieving either one. In this paper, we propose a partial direct-mapped die-stacked DRAM cache to achieve the both objectives simultaneously, called \textit{Gemini}, which is motivated by the following observations: applying unified mapping policy to different blocks cannot achieve high cache hit rate and low hit latency in terms of mapping structure. \textit{Gemini} cache classifies data into leading blocks and following blocks, and places them with static mapping and dynamic mapping respectively in a unified set-associative structure. \textit{Gemini} also designs a replacement policy to balance the different blocks miss penalty and the recency, and provides strategies to mitigate cache thrashing due to block type transitions. Experimental results demonstrate that \textit{Gemini} cache can narrow the hit latency gap with direct-mapped cache significantly, from 1.75X to 1.22X on average, and can achieve comparable hit rate with set-associative cache. Compared with the state-of-the-art baselines, i.e.,  enhanced Loh-Hill cache, \textit{Gemini} improves the IPC by up to  20\% respectively.

\end{abstract}

%
%


\maketitle

\section{Introduction}
\label{sec_introduction}


The 3D die-stacking DRAM provides high bandwidth, low latency and large capacity, mitigating memory wall. Due to its gigascale capacity, the 3D die-stacking DRAM could not replace off-chip DRAM and has been proposed to be architected as the last level cache, referred to as DRAM cache ~\cite{loh11,loh12,qureshi12, jevdjic13, chou16}. The 3D die-stacking DRAM is  helpful for a lot of applications, such as peer-to-peer live streaming ~\cite{liao}.However the DRAM cache's tag storage overhead caused by the large capacity of 3D DRAM makes it challenging to design high performance DRAM cache. For example, with 512 MB DRAM cache, its tag lines storage overhead is about 24 MB with 64B cache line. In order to address this issue, prior research work proposes two solutions: 1) storing tags in DRAM cache with small granularity of cache line,  and 2) storing tags in SRAM with large granularity of cache line. However, they still have limitations. Co-locating data lines and tag lines in DRAM cache serializes the tag access and data accesses from DRAM cache, increasing cache hit latency. Large size cache line suffers from large bandwidth overhead, DRAM cache capacity scalability, and under-utilization of space.  This paper focuses on the DRAM cache design with smaller cache line size.

Hit rate and hit latency are the two important performance metrics in DRAM-d cache designs. The Alloy Cache~\cite{qureshi12} was proposed to merge a data line with its tag in tag-and-data unit(TAD) and performs tag look up by issuing a CAS command to stream out a TAD. In this way, the tag-then-data serilization has been elimiated, reducing hit latency. However, TAD restricts cache organization to be direct-mapped cache and suffers from lower hit rate.  LH Cache ~ \cite{loh11,loh12} architects the DRAM cache as set-associative cache  by co-locating the tags with data blocks in the same row and achieving high hit rate. On serving a request, the DRAM cache controller needs to retrieve all tag lines belonged to a set by issuing a CAS DRAM command and multiple bus bursts before determining the location of the requested data line. This tag access latency increases the hit latency.  Ref. ~\cite{Huang2014,Hameed14} proposed to cache tags in a small on-chip SRAM to speed up the tag lookup for the set-associative DRAM cache. On tag-cache hit, the data block can be fetched from DRAM cache without accessing the tag in DRAM cache. However, on tag cache miss, the tag is fetched to the tag cache before the data block is accessed. Therefore, the tag-then-data access serialization can not  be completely removed from data access path by using tag cache, resulting sub-optimal performance. These research work on  the DRAM cache as  the directed-map cache and set associative cache exclusively and failed to optimize the hit latency and hit rate simultaneously.

We have made some interesting observations on the set associative cache with tag cache in SRAM. On the tag cache miss, a batch of tags will be fetched from the DRAM cache into the SRAM tag cache before the requested data is accessed in DRAM cache. The target data block in the first access of this set is refereed to as the leading block, and the rest are referred to following blocks. Our experimental results demonstrate that the individual blocks exhibit the stable block type, which is either a leading block or following block. Furthermore, we observed that it is the leading block that incurs the tag fetching overhead, while the following blocks can benefit from the fast tag lookup in SRAM. Our experimental results on 18 workloads (Section 5) show that on average  89\% tag fetching are triggered by the leading blocks, and these tag fetching increase the leading block's hit latency by 1.7X-2.3X compared with the direct-mapping cache. If  applying direct mapping to leading blocks, their hit latency could be significantly reduced. In addition,  most following blocks hit tag cache and over 97\% of their hit latency is caused by the data fetching from DRAM cache, enjoying the free ride provided by the leading blocks tag fetching. 

Motivated by the above key observations, we propose \textit{Gemini}, a partial direct-mapped DRAM cache that exploits block differentiation with the hybrid mapping policy to achieve low hit latency and high hit rate simultaneously. Specifically, we apply static mapping to leading blocks to reduce cache hit latency, and dynamic mapping for following blocks with hit rate degradation. \textit{Gemini}faces two challenging issues. First, we find that leading blocks and following blocks have different miss penalty in terms of latency and bandwidth. For example, a leading block miss incurs a latency of 273 cycles, which is 1.3X higher than that of a following block. This phenomenon should be considered in the design of cache replacement policy. Second, the frequent block type transitions are needed to handled to avoid hurting performance in some workloads. 


The main contributions are summarized as follows.

\begin{itemize}
	\item We observe that  data blocks are classified to be leading blocks and following blocks and the individual data block possess the stable block type. Furthermore, two types of data blocks have distinct impacts on the hit latency and hit rate of DRAM cache mapping policies.  
	\item We propose a partial direct-mapped DRAM cache, called \textit{Gemini}, which applies static and dynamic mapping to leading and following blocks, respectively, to achieve low hit latency and high hit rate simultaneously. In addition to the novel mapping scheme, we also propose a cache replacement policy called Range-Variable CLOCK (RV-CLOCK) considering the different miss penalties for data blocks with different block type. Furthermore, to deal with the frequent block type transitions, we propose the priority reservation mechanism with a high frequency variation filter.
	\item Through extensive evaluations, we demonstrate that \textit{Gemini} cache can narrow the hit latency gap with direct-mapped cache significantly, from 1.75X to 1.22X on average, and can achieve comparable hit rate with set-associative cache. Compared with the state-of-the-art baselines, i.e., Loh-Hill cache enhanced with tag cache, \textit{Gemini} improves the IPC by up to 20\% respectively.
\end{itemize}

The rest of this paper is organized as follows. we present the background in Section 2 and motivations in Section 3. Section 4 introduces system design. Experimental methodologies are given in Section 5, followed by evaluations in Section 6. Related works are summarized in Section 7. Section 8 concludes this paper.

\section{background}
\label{sec_2_background}


\subsection{DRAM Cache Organizations}
\label{sec_2_1_cache_organizations}

\begin{figure}
	\centering
	\includegraphics[width=0.475\textwidth,height=1.2in]{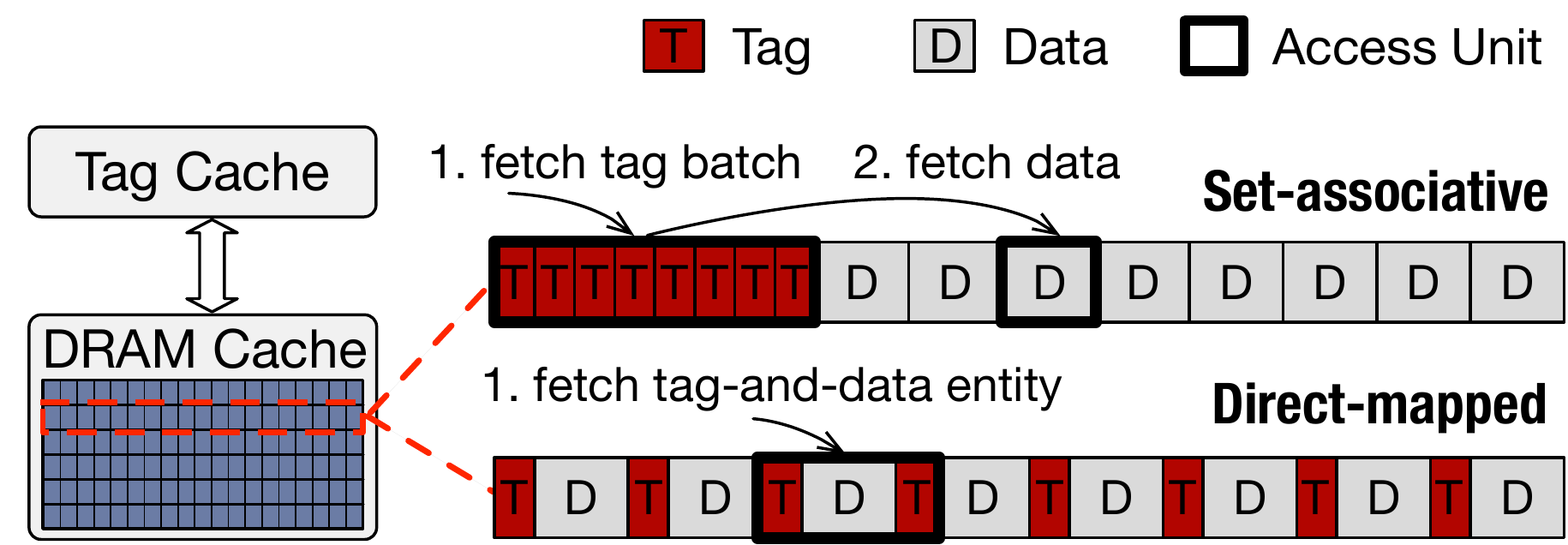}
	\caption{\normalsize{\textbf{basic organizations of set-associative cache and direct-mapped cache}}}
	\label{fig_two_basic_cache}
\end{figure}

Similar to the conventional SRAM cache, DRAM cache has tag and data for each data block. Since the large DRAM cache makes it impractical to accommodate the correspondingly large tag in SRAM, LH Cache proposes to store the tag and data in the DRAM cache. LH Cache architects DRAM cache as a set associated cache by storing tag and data of a set in one DRAM row. The set-associative can reduce the conflict misses and benefit system performance. To service a request, the DRAM cache controller checks the tag and then reads the data line according to the outcome of tag query, shown in the Fig.~\ref{fig_two_basic_cache}, increasing the hit latency. This serialization of tag access and data line access increases the hit latency in DRAM cache, resulting in sub-optimal performance. In order to address this issue, the Alloy Cache proposes to trade the low hit latency for the low hit rate. The Alloy cache organizes the DRAM cache as a direct-mapped cache and combines a data line with its tag line, referred to as tag-and-data units(TAD). This merging of data line and tag line removes the searching correct way from the data access path and directly accesses the TAD to avoid the serialization of tag and data access depiected Fig.~\ref{fig_two_basic_cache}. However, it suffers from the low hit rate because of the direct-mapped cache organization.  Caching tags in small size on-die SRAM ~\cite{jevdjic13, chou15, gulur14} was proposed to mitigate the issue of tag-then-data serialization. On a tag miss, the tags of a set are fetched in batch to the on-chip SRAM. Due to the spatial locality, the later accesses to the same set result in  tag cache hit and the DRAM cache can directly access the data in the DRAM cache, without probing tag in the DRAM cache.

\subsection{Access Latency Breakdown}
\label{sec_2_2_latency}

\begin{figure}
	\centering
	\includegraphics[width=0.49\textwidth,height=2.5in]{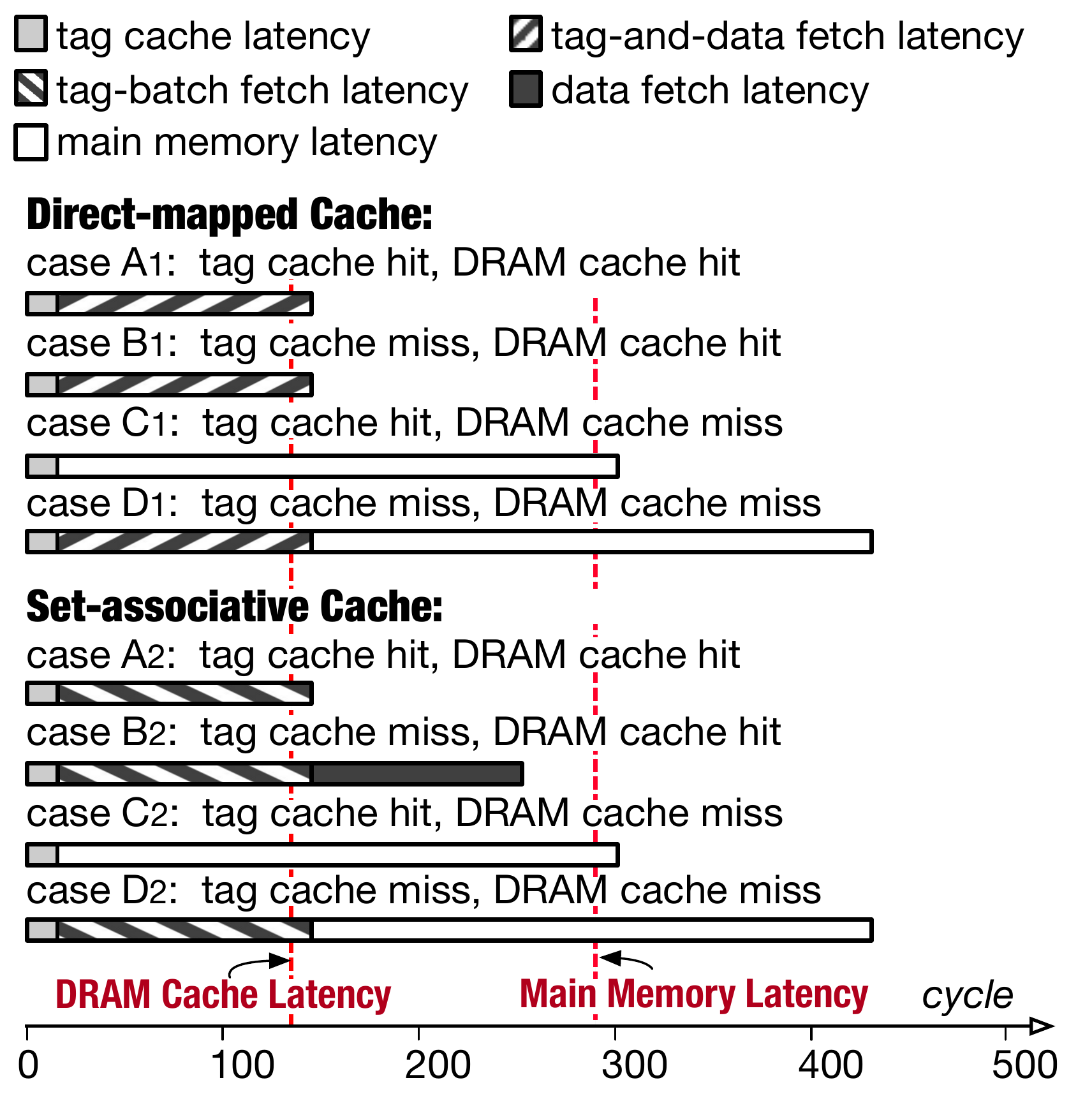}
	\caption{\normalsize{\textbf{Latency breakdown of direct-mapped cache and set-associative cache}}}
	\label{fig_latency_breakdown}
\end{figure}


Figure~\ref{fig_latency_breakdown} illustrates the access latency of the set associative DRAM cache  and directed-mapped DRAM cache. The direct-mapped cache can offer the lowest hit latency by fetching tag and data with a single request on DRAM cache hit (cases $A_1$ and $B_1$). However, if the tag cache and the DRAM cache are both missed (case $D_1$), an extra cache probe must be performed, before accessing off-chip memory. If the tag cache indicates that the data block is not present in DRAM cache (case $C_1$), the request will be directly sent to off-chip memory. On tag cache miss, the set-associative cache suffers the cache probe latency even if the request can hit DRAM cache (case $B_2$). For the rest of the cases (cases $A_2$, $C_2$, and $D_2$), the set-associative cache acts in the same way as the direct-mapped cache, thus have similar access latency. The benefits of the direct-mapped structure lies in low DRAM cache hit latency, but the performance is sensitive to cache hit rate. The tag fetching  introduced  in the set-associative cache makes the DRAM cache hit  latency close to the off-chip memory, negating the benefit of DRAM cache, as show in Figure~\ref{fig_latency_breakdown}.



\section{motivation}
\label{sec_3_motivation}
\subsection{ Leading Blocks and Following Blocks}

\begin{figure}
	\centering
	\includegraphics[width=0.52\textwidth,height=2.1in]{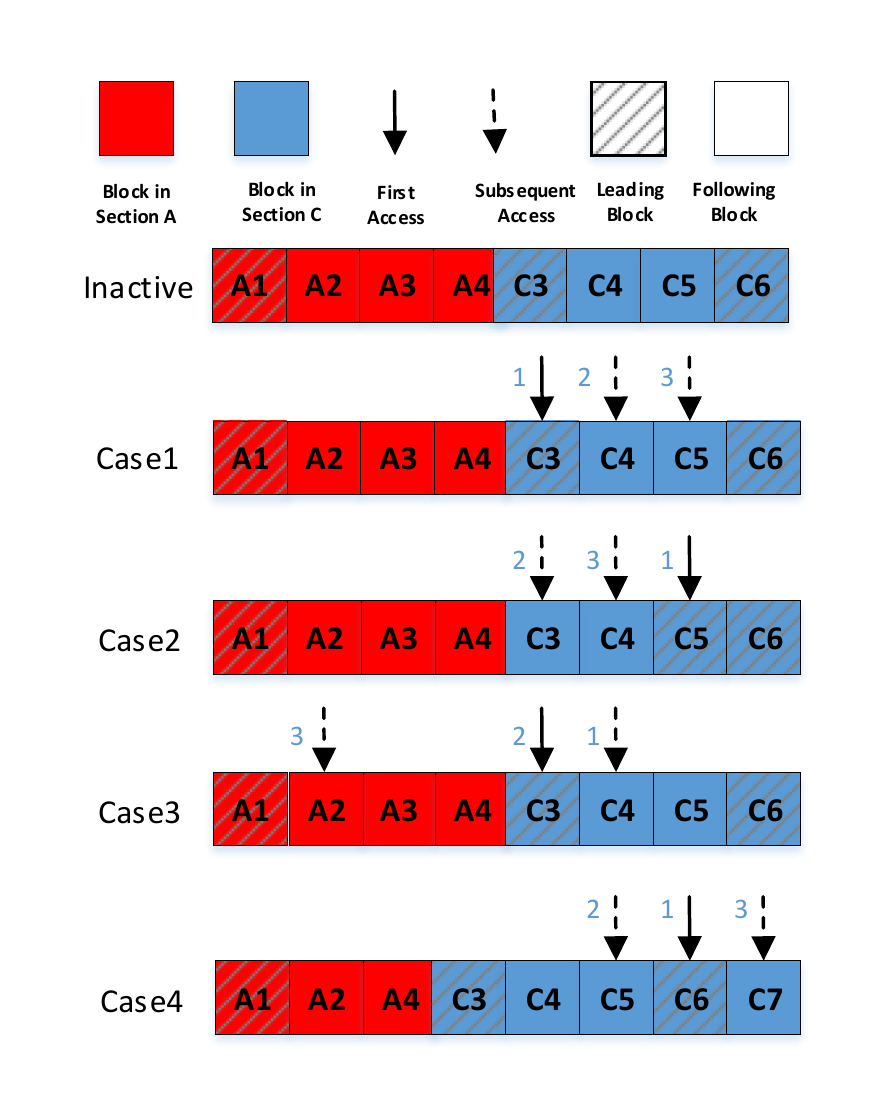}
	\caption{\normalsize{\textbf{Leading and Following Data Blocks in a Set}}}
	\label{fig_leading_following}
\end{figure}

We first define several  terminologies that will be used in the rest of this paper. A \textit{section} is defined as a continuous logical address region mapped to a single cache set and hence data blocks in the same region are mapped to the same cache set. This mapping enable us to preserve the spatial locality exhibited in workloads. There could be multiple sections mapped to the same set. For example, the data blocks in section A and section B reside in the same set, shown in the Fig.~\ref{fig_leading_following}. Although this mapping is inferior to the conventional mapping in high level caches, our experimental results show it achieves the hit rate  close to the conventional mapping in of DRAM cache, due to the weak locality.
Each L3  cache miss  checks the tag cache before proceeding to DRAM cache. On a tag cache miss, all tags in the target DRAM cache set, referred to as tag batch, are fetched to the tag cache in SRAM. A \textit{section} is active if any of its tags is cached in the tag cache, or is currently being fetched. Otherwise, it is inactive. A \textit{leading block} is the  block to make  the corresponding \textit{section} to transition from inactive state to active state, and the \textit{following blocks} are ones accessed later until the \textit{section} becomes inactive, which is caused by tag cache replacement. In other words, a leading block is the data block experiencing tag cache miss and the subsequently accessed blocks in the same section are following blocks when their tags are in the tag cache. 

We use a simplified example to explain the leading blocks and following blocks shown in Fig.~\ref{fig_leading_following}.  Assume that there are eight data blocks from the section A and the section C in the same DRAM cache set, and their tags are not in tag cache. Accordingly, these two sections are in inactive state. We further assume section A has one leading block A1 and section C has two leading blocks C3 and C6. At the case 1 with the access sequence of C3, C4 and C5, the  C3 is a recurring leading block, because it misses the tag cache. After the C3 being served, the set's tag has been fetched to the tags cache, and the subsequent accesses to C4 and C5 hit tag cache. They are following blocks. The case 2 shows the new leading block, C5, in the section C because its access misses the tag cache, explaining the existence of multiple leading blocks for a section. At the case 3,  the  block A2 is a following block because the block C4  has activated the section A.  The case 4 presents that the block C7 is also a following block even its tag is not in tag cache due to DRAM cache miss. This is because its section C has been activated by the C6.

%


\subsection{Block Type Stability}
\label{sec_3_4_block_type_consistency}

\begin{figure}
	\centering
	\includegraphics[width=0.47\textwidth,height=1.5in]{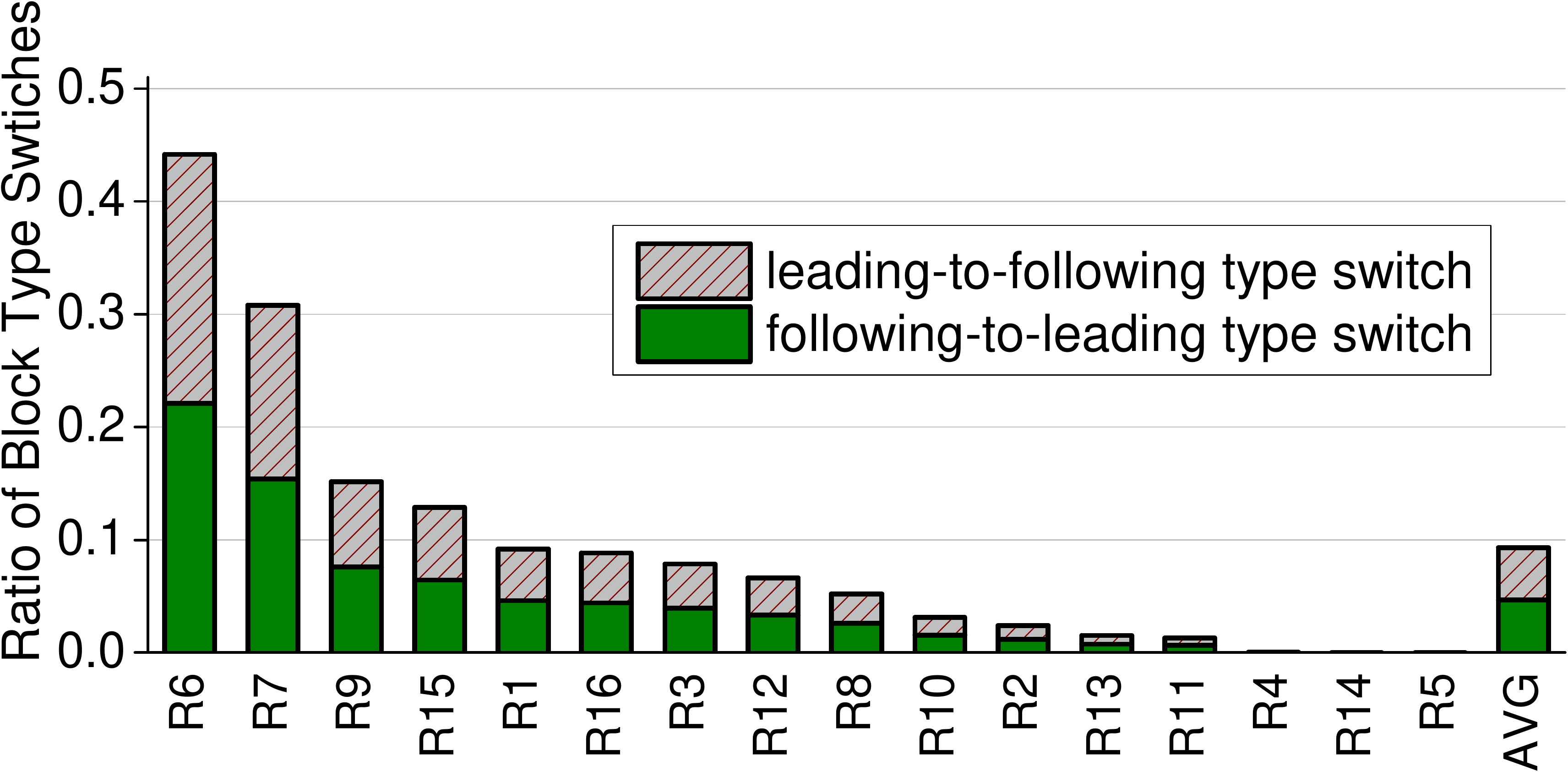}
	\caption{\normalsize{\textbf{The ratio of block type switches}}}
	\label{fig_block_type_consistency}
\end{figure}

We measure the ratio of blocks that have data type transitions for 10 workloads and they are  decreasingly  ordered by the transition ratio shown in Figure~\ref{fig_block_type_consistency}. The average ratio of block type switches is less than 0.05. These results demonstrate that block types are almost stable. The reason why blocks have stable type is that each section contains data objects accessed by the specific code and the inherent semantic
of the code manipulates these data objects with its unique pattern . For example, a field variable \textit{var} in a struct is accessed by the specific function before other field variables in the struct variable, which is most likely in the same region. Next time, the execution of this function repeats the same access pattern on the struct variable. In this case, the data block containing the variable \textit{var}  is a stable leading block. We are going to exploit the block type stability to optimize DRAM cache design.

It is worth to note that the workload R7 and R6 have non-negligible data type switches and these unstable blocks make our proposed optimization ineffective. We will discuss how to address this issue at the  next section.


\subsection{Impacts of  Leading Blocks and Following Blocks  }
\label{sec_3_1_blocks}


Set-associative cache offers high hit rate, but incurs large hit latency. In the DRAM cache, the large associativty necessitates the
tags co-located with the data blocks in the same row and tags are accessed before data blocks. Such serialization of tag and data blocks is the root reason for the large hit latency. 

Tag cache in SRAM is an effective way to reduce the hit latency in DRAM cache. Upon a tag cache miss, all tags of the set are fetched from the DRAM cache and then are stored in SRAM to accelerate the tag lookup for following accesses. Due to spatial locality, most of following accesses benefit from the quick tag lookup in SRAM, avoiding tag fetching from DRAM cache, because the leading blocks take the responsibility to fetch tags to tag cache. 

\begin{figure}
	\centering
	\includegraphics[width=0.47\textwidth,height=1.0in]{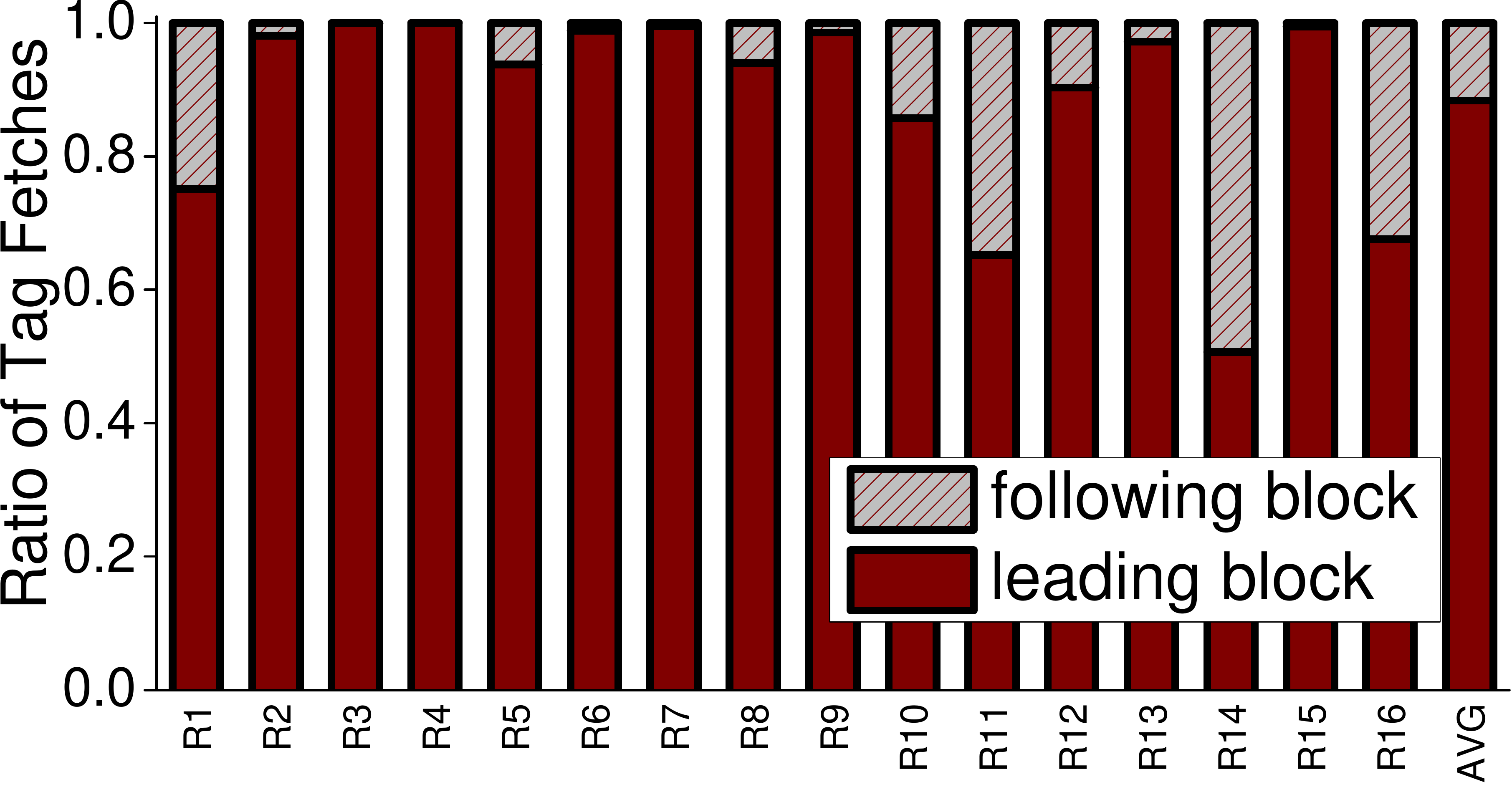}
	\caption{\normalsize{\textbf{Distributions of tag fetches}}}
	\label{fig_tag_fetch_distribution}
\end{figure}

Our experiments show the proportion of tag fetching caused by the leading blocks and the following blocks in Figure~\ref{fig_tag_fetch_distribution} for 10 workloads (details in Table~\ref{tbl_workloads}). We find that on average 89\%  tag fetches are triggered by the leading blocks. Therefore, 89\% the serializations of  data and tags accesses are caused by the leading blocks in the setting of tag cache and hence removing these serializations could effectively reduce the hit latency. This observation motivates us to apply the direct mapping to the leading blocks.  Without tag probation latency, the direct mapping can remove  89\% tag-then-data serializations, reducing the hit latency.  


Differing from leading blocks, most following blocks involve only data fetching from DRAM cache since their tags are cached in SRAM. Our experimental results show that 97\% of following block hit latency is caused by the data block fetching. Therefore, direct mapping  can hardly further reduce hit latency for following blocks. Moreover, direct mapping seriously hurts hit rate. The above observations motivate us to apply dynamic mapping for following blocks to achieve higher hit rate and smaller hit latency.



\subsection{Block Miss Penalties}
\label{sec_3_3_replacement_costs}

\begin{figure}
	\centering
	\includegraphics[width=0.47\textwidth,height=1.2in]{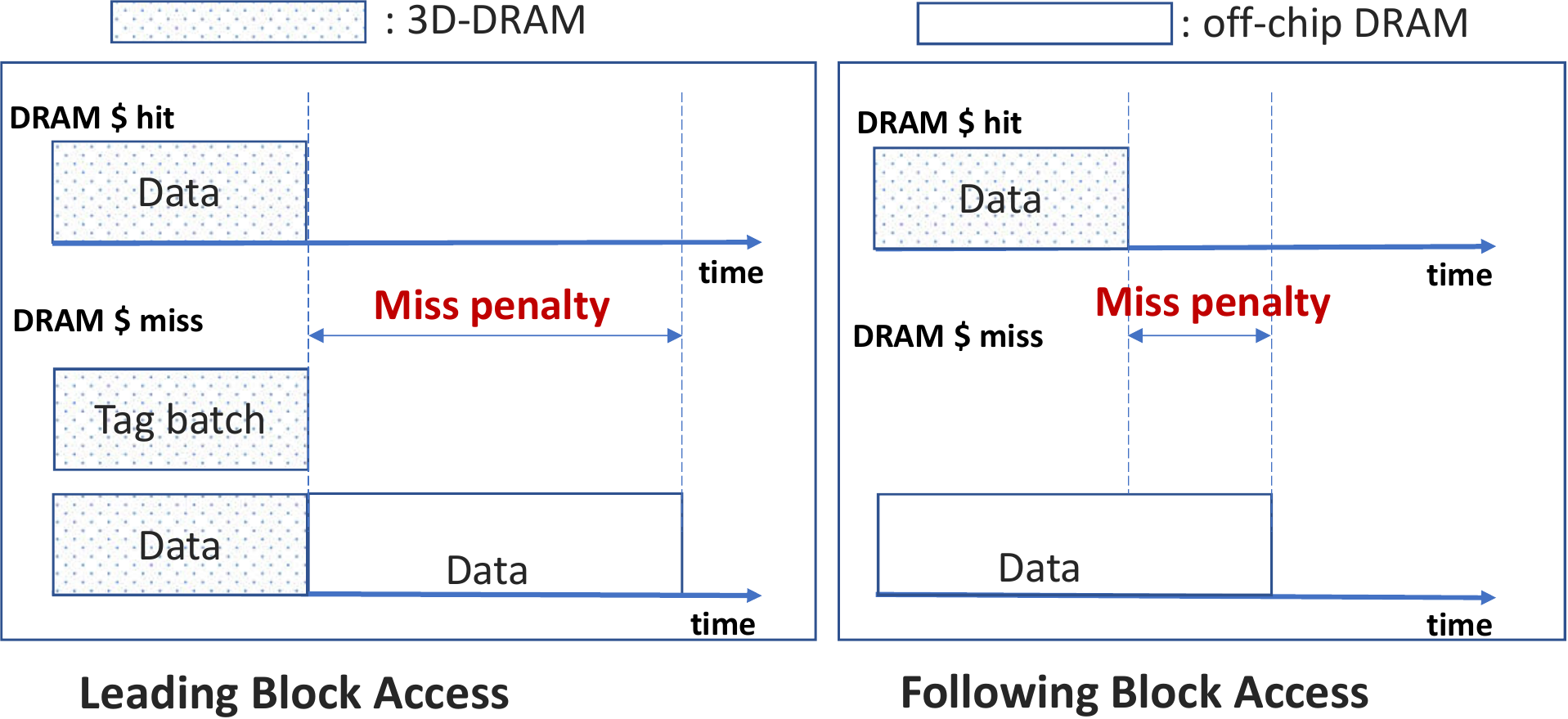}
	\caption{Miss Penalty Comparison between Leading Block and Following Block}
	\label{fig_miss_penalties}
\end{figure}

The Fig.~\ref{fig_miss_penalties} compares the miss penalty of leading block with following block. The left half of the Fig.~\ref{fig_miss_penalties} shows the DRAM cache hit and miss for the leading block. Since the leading block is statically mapped to
the DRAM cache, the DRAM cache controller directly fetch the requested leading block from the DRAM cache, requiring one 3D DRAM access.  After reading the leading block from DRAM cache, the cache controller detects the cache miss and fetches the tag batch from DRAM cache and the requested data block from off-chip DRAM simultaneously, shown in the bottom of the Fig.\ref{fig_miss_penalties}.
This figure clearly shows the miss penalties of the  leading block are a data block access from off-chip DRAM in term of latency. The right half of Fig.~\ref{fig_miss_penalties} shows miss penalty for the following block is the access latency difference between the DRAM cache and off-chip DRAM. For the following block, the cache controller can quickly determine the cache miss/hit with the help of the tag batch in the SRAM. Upon hit, the requested block is fetched from the DRAM cache. Otherwise, one off-chip DRAM access is involved. Therefore, the leading block has larger miss penalties than following block in term of both latency and bandwidth consumption.

Our experimental results support the above analysis. Table~\ref{tbl_replacement_costs} shows the average latency, as well as bandwidth  for both the  leading block and the following block on DRAM cache hit/miss under 14 workloads. We find that the leading block miss penalty is 273 cycles with  a  64-byte data transfer over off-chip DRAM, and the following block miss penalty is 112 cycles. Such  miss penalty differences between  leading blocks and following blocks motivates us to design a DRAM cache replacement policy to minimize leading block misses.

\begin{table}
	\caption{\normalsize{\textbf{The miss penalty of leading blocks and following blocks.}}}
	\small
	{\centering \ra{1.1}
		\begin{tabular}{@{}L{1.8cm}|C{0.4cm}|C{1.5cm}|L{2.75cm}@{}}
			\toprule
			\multicolumn{1}{c}{} & \multicolumn{1}{c}{} & \multicolumn{1}{l}{$\mathbf{Latency_{(cyc)}}$}  & \multicolumn{1}{l}{$\mathbf{Bandwidth_{(byte)}}$} \\
			\hline
			\hline
			\multirow{4}{*}{Leading block} & \multicolumn{1}{c|}{$Hit$}  & 185 & Cache: 128$_{(tag+data)}$\\
			\cline{2-4}
			& \multirow{2}{*}{${\ }{\ }{\ }Miss$} & \multirow{2}{*}{458} & Cache: 128$_{(tag+data)}$\\
			& & & Memory: 64$_{(data)}$\\
			\cline{2-4}
			& \multicolumn{1}{c|}{$Penalty$}  & \textbf{273} & \textbf{Cache: 64$_{(data)}$} \\
			\hline
			\hline
			\multirow{3}{*}{Following block} & \multicolumn{1}{c|}{$Hit$}  & 147 & Cache: 64$_{(data)}$ \\
			\cline{2-4}
			& \multicolumn{1}{c|}{$Miss$}  & 259 & Memory: 64$_{(data)}$ \\
			\cline{2-4}
			& \multicolumn{1}{c|}{$Penalty$}  & \textbf{112} & \textbf{--} \\
			\hline
		\end{tabular}
	}
	\label{tbl_replacement_costs}
\end{table}

%
%
%

\section{design}
\label{sec_4_design}


\subsection{Cache Organization}
\label{sec_4_1_cache_organization}

\begin{figure}
	\centering
	\includegraphics[width=0.49\textwidth, height=2.7in]{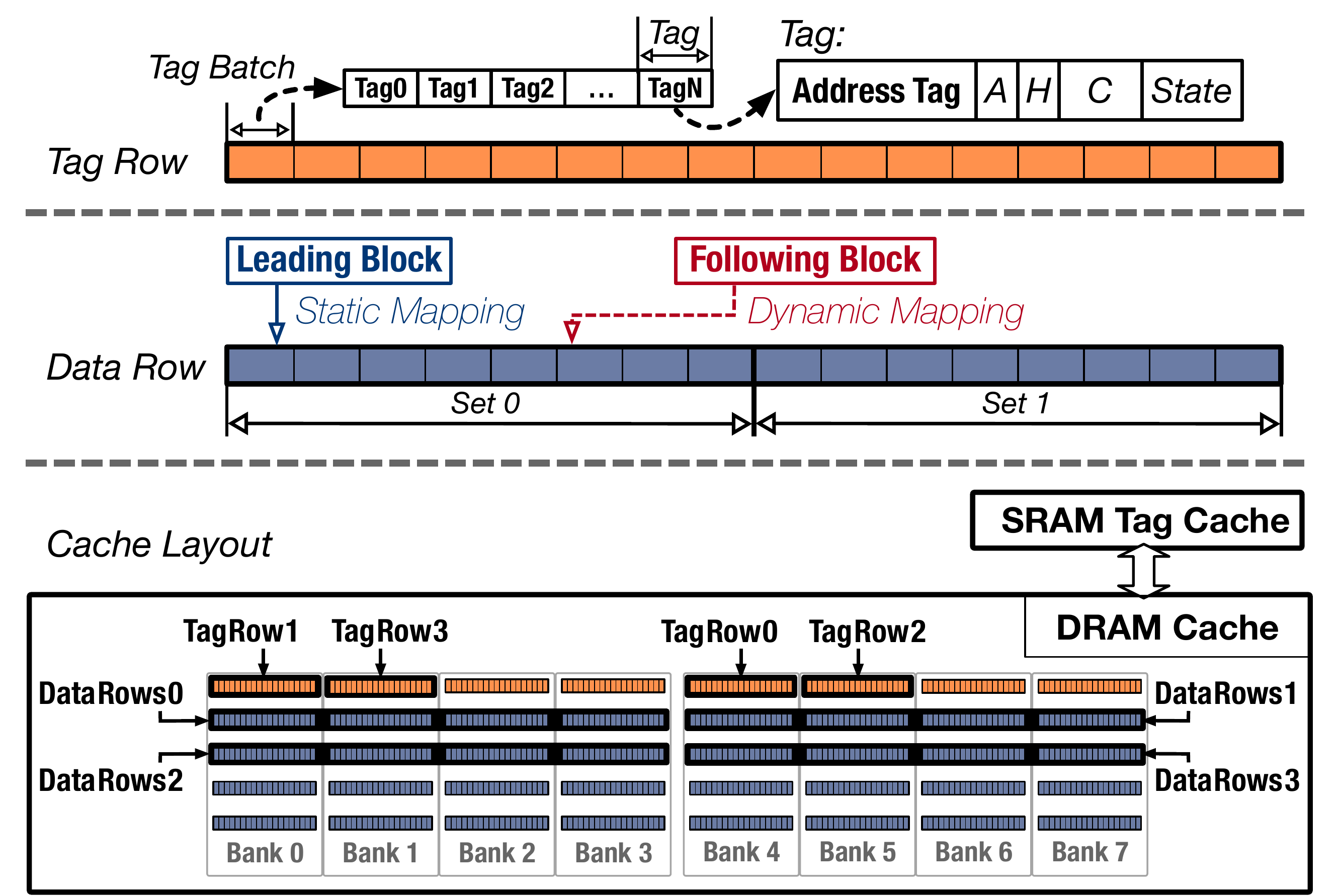}
	\caption{\normalsize{\textbf{Organizations of \textit{Gemini} cache}}}
	\label{fig_gemini_organization}
\end{figure}

As discussed in section~\ref{sec_3_motivation}, the application of the different mappings to leading blocks and following blocks could  improve the hit latency and hit rate simultaneously in the context of tag cache. This is the  key idea of the \textit{Gemini}. The type of the requested data blocks determines how the \textit{Gemini} serves the request.  For the  leading block, its direct mapping  enables cache controller to fetch the data from the known position in the DRAM cache set without consulting the tag and the related tag batch is also transferred to the tag cache in SRAM.  In this way, the cache controller can detect the DRAM cache miss for the leading block, after checking the tag batch.  In order to concurrently transfer both leading block and tag batch, the data set and the corresponding tags are stored in different banks, shown in Figure~\ref{fig_gemini_organization}. For the following block, the requested data needs to consulting its tag to determine the position in the set, since the dynamic mapping is applied to the following block. The following block is most likely to hit the tag cache and is quickly streamed out from DRAM cache.  


The organization of tag row is  depecited at the top of Figure~\ref{fig_gemini_organization}. A tag batch contains the tags for the data blocks stored in the corresponding cache set, which is similar to the tag batching mechanism in the previous work\cite{loh11}. Four extra bits are introduced to a tag: reference bit (\textit{A}), priority bit (\textit{H}), and a two-bits filter (\textit{C}). The reference bit indicates the recency of the block. The priority bit denotes the block's caching priority and also acts as the block type indicator (1 for leading and 0 for following). The filter bits  monitor the transitions of block type  discussed in Section~\ref{sec_4_3_type_switch}.

The data row's organization is illustrated in the middle of Figure~\ref{fig_gemini_organization}. In conventional designs, the direct-mapped cache and the set-associative cache are commonly viewed as two exclusive structures. Generally, the cache associativity and the mapping policy are closely coupled: 1-way set is the direct mapping, and n-way set is referred to as the dynamic mapping. In order to support both static mapping and dynamic mapping in an unified structure, we decouple the set associativity from the data mapping policy. Gemini is basically organized in a set-associative manner.  Each DRAM row has several sets and each set contains 16 data lines. Within each cache set, data blocks can be mapped via either static or dynamic mapping, depending on their block types, as shown in the Figure~\ref{fig_gemini_organization}. Specifically, following blocks are installed in the set according to the replacement algorithm, while leading blocks' positions at the set are determined by the static mapping, which is the predetermined hash function of the block address.

The layout of Gemini cache is demonstrated at the bottom of Figure~\ref{fig_gemini_organization}. The DRAM rows are divided into tag rows and data rows. Tags and data are mapped to the rows of different banks to improve the bank level parallelism. The ratio between tag rows and data rows equals to $tag\_size/data\_size$. In our design, $tag\_size=4bytes$, $data\_size=64bytes$. For illustrative purpose, we show a tag row coupling with four data rows, while  the ratio is $1:16$ in real case .

\subsection{Mapping Policy}
\label{sec_4_2_mapping_policy}

A requested block is a currently leading block if it access the inactive section whose tags are not in tag cache at the time point. Otherwise, it is a currently following block. We apply static address mapping and dynamic address mapping to leading blocks and following blocks  respectively to determine the position in the cache set. The tag of the data block has the priority bit to indicate the block type. For the leading block, the address of the request block is hashed on the predetermined function and its outcome is the position for this leading block. After getting position of the leading block, the memory controller directly retrieves the data block from the DRAM cache. It is possible that there are multiple leading blocks in the one cache set since the position depends on the request block address. In that case, position conflicts caused by the  hash function could lead to eviction. For the following block, the position in the set can be determined by the probing the tags stored in the tag cache in the case of tag cache hit. Otherwise, the cache replacement algorithm determines the position to be installed in the cache set. As shown in algorithm~\ref{alg_basic_type_switching}, we use the $GetPosition( )$ to denote the position determined by the cache replacement policy. 

Data type transitions are needed to be handled. The  algorithm~\ref{alg_basic_type_switching} shows how the \textit{Gemini} handle data type transitions. For the transition to the following one, \textit{Gemini} resets its priority bit (\textit{line 4,5}). For the transition to the  leading one, \textit{Gemini} tries to migrate it to the statically mapped position (\textit{line 6-17}). If the data block's current position equals to the given  static mapping (\textit{line 7-8}), no migration is needed , and \textit{Gemini} sets its priority bit to 1. Otherwise, i.e., two positions are not equal, one of the following actions is taken: 1) If the statically mapped position is  high priority, and is recently referenced, \textit{Gemini}  just resets its reference bit (\textit{line 10,11}), without migration. 2) In the rest cases, \textit{Gemini} migrates the data  to its statically mapped position, with the expectation that the data will be continually accessed as a leading block in the near future.

\begin{algorithm}
	\caption{{{Mapping Policy}}}
	\label{alg_basic_type_switching}
	\KwIn{\textit{block\_address, type} }
	
	\textit{position}$\gets$GetPosition(\textit{block\_address})\\
	\textit{static\_position}$\gets$StaticMapping(\textit{block\_address})\\
	\textit{last\_type}$\gets$GetType(\textit{block\_address})\\
	\If{last\_type{\ }$=${\ }Leading \textbf{\textup{and}} type{\ }$=${\ }Following}{
		\textit{cache[position].priority\_bit}$ {\gets}$\textit{0}\\
	}
	\If{last\_type{\ }$=${\ }Following \textbf{\textup{and}} type{\ }$=${\ }Leading}{
		\uIf{position{\ }$=${\ }static\_position}{
			\textit{cache[position].priority\_bit}$ {\gets}$\textit{1}\\
		}\Else{
			\uIf{cache[static\_position].type{\ }$=${\ }Leading \textbf{\textup{and}} cache[static\_position].reference\_bit{\ }$=${\ }1}{
				\textit{cache[static\_position].reference\_bit}$ {\gets}$\textit{0}\\
			}\Else{
				\If{IsDirty(cache[static\_position])}{
					WriteBack(\textit{cache[static\_position]})\\
				}
				\textit{cache[static\_position]}$ {\gets}$\textit{cache[position]}\\
				Free(\textit{cache[position]})\\
				\textit{cache[static\_position].priority\_bit}$ {\gets}$\textit{1}\\
			}			
		}
	}
\end{algorithm}

\subsection{Replacement Policy}
\label{sec_4_2_replacement_policy}

As discussed in section~\ref{sec_3_motivation}, leading blocks incurs more DRAM cache miss penalty than following block in terms of latency and bandwidth.  The cache replacement algorithm can improve DRAM cache performance by exploiting this observation. The replacement algorithm assigns high caching priority to leading blocks  and low priority to following blocks. To this end,  the priority bit for each data block is introduced. The priority is set to to be 1 and 0 for the leading blocks and the following blocks respectively. In this way, the replacement algorithm attempts to keep the leading blocks in cache longer to amortize its miss penalty. However, the leading blocks can gradually encroach the cache space of the following blocks due to the high priority. This  intensifies the cache space competition with the following blocks. On the other hand, the cache space occupied by the cold leading blocks lead to lower cache utilization without considering the overall cache hotness.

To balance the miss penalty and cache hotness, we propose the Range-Variable CLOCK (RV-CLOCK) algorithm, which is based on the CLOCK algorithm due to its close approximation of LRU \cite{jiang2005clock} with low overhead.  In the CLOCK algorithm , a reference bit indicates the recency of a data block. If the reference bits of all following blocks have been set, these recently accessed following blocks are chosen as victims for the later arriving following blocks, and the cache thrashing occurs for the following blocks, hurting performance. In order to address this issue, we attempt to evict a cold leading block to increase cache allocation of following blocks. Specifically, we run the CLOCK in the scope of the whole cache set, overriding the priority settings.

\begin{figure}
	\centering
	\shortstack{
		\vspace{3pt}
		\includegraphics[width=0.45\textwidth,height=0.4in]{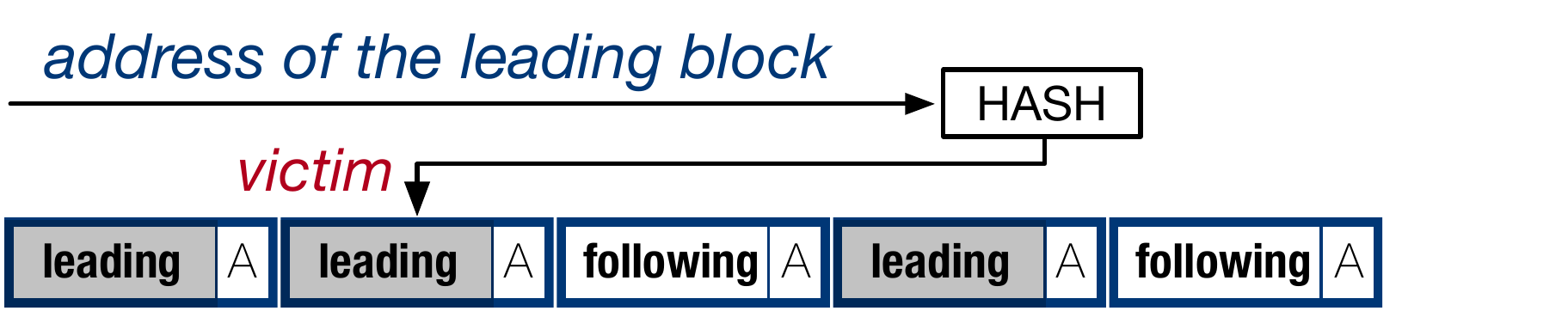}\\
		\vspace{6pt}
		{1. The victim selection upon leading block insertion}
		\label{fig_leading_replacement}
	}
	\shortstack{
		\vspace{-6pt}
		\includegraphics[width=0.45\textwidth, height=0.8in]{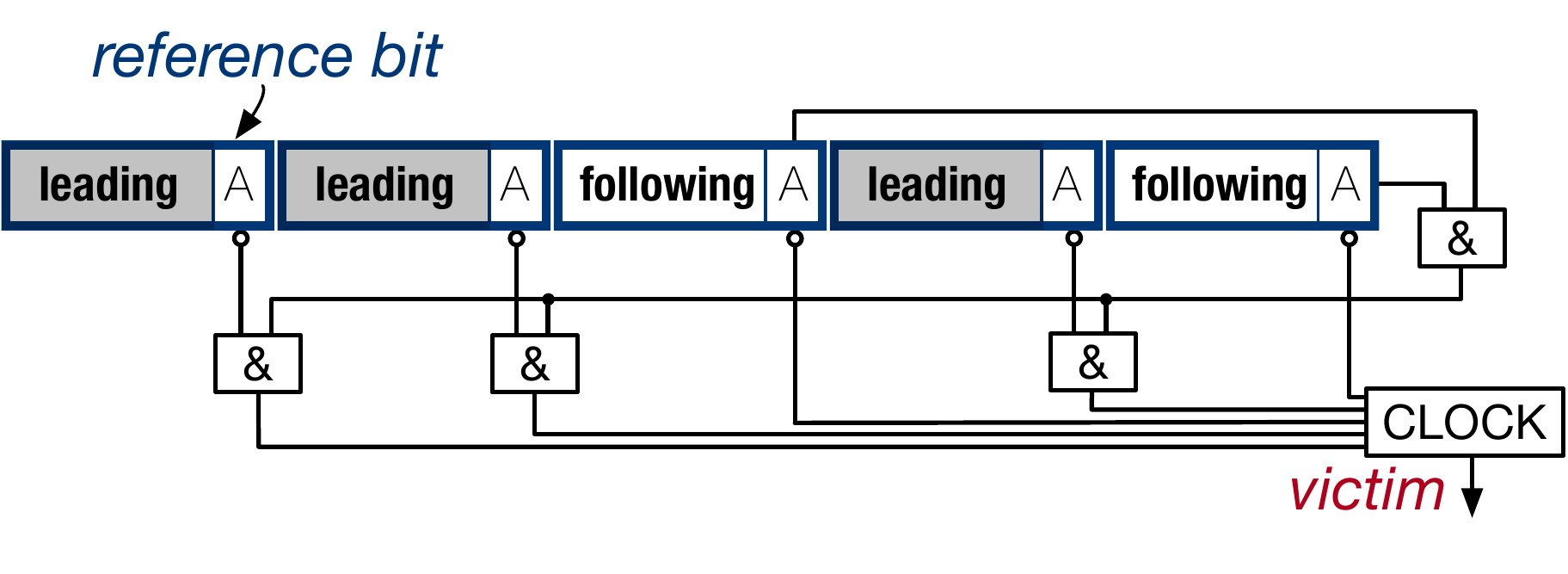}\\
		{2. The victim selection upon following block insertion}
		\label{fig_following_replacement}
		\vspace{-6pt}
	}
	\caption{\normalsize{\textbf{The victim selection policies}}}
	\label{fig_replacement_policy}
\end{figure}

This algorithm can be implemented with low hardware cost. As illustrated in Figure~\ref{fig_replacement_policy}, the following blocks' reference bits act as a mask of the CLOCK's working range. If the reference bits of following blocks are all set, the algorithm runs the CLOCK for all lines to choose the victim. Otherwise, the leading blocks are masked, and thus can not be evicted.

\subsection{Managing Type Switching}
\label{sec_4_3_type_switch}



The mapping policy works well for the data with stable block types, which is common discussed in the section~\ref{sec_3_4_block_type_consistency}. However, there exist some workloads whose block type transitions frequently occur. Such transitions lead to larger number of data block migrations, degrading performance. Furthermore, the mapping policy mistakenly downgrades the blocks' priority and prematurely evicts blocks. For example, if an unstable block switches from leading block to following block, the mapping policy reset its priority bit, increasing the block's eviction probability.  At the next active period of the \textit{section} , it is highly likely that this block will switch back to be a leading block. In such a case, the mapping policy untimely evicted this leading block, suffering the costly miss penalty.


To address the issue, we further study the block type stability. For a single block,  the $L_{stable}$  represents the number of occurrences of consecutive leading or following type. For example, consider a type sequence of \{$leading$ $\to$ $leading$ $\to$ $following$ $\to$ $following$ $\to$ $leading$\}. Their  $L_{stable}$ are 2, 2, 1. The larger the value of $L_{stable}$ is, the more stable the block remains the same type. Our experimental results show the $L_{stable}$ are not greater than 10. After examining the workload whose percentage of block type transitions is larger than 10\%, we find that 88\% of  $L_{stable}$ is not larger than 2. This implies that the type of these blocks  tend to continuously change. 

Motivated by the above observation, we propose a priority reservation mechanism with a \textit{frequent type transition filter} to handle unstable blocks. The key idea is that we track the blocks whose types frequently change, and assign high priority to them to avoid the potential miss penalty on their $following$ $\to$ $leading$  transitions. As discussed early, we mainly focus on the cases with $L_{stable}\leq 2$. Specifically, we use a two-bits counter to filter out these blocks. If there is a $following$ $\to$ $leading$ transition, the block's counter is increased by 2. The counter is decreased by 1 for the $following$ $\to$ $following$ transition. After reaching 0, the two-bits counter remains "00'' state for the $following$ $\to$ $following$ transitions. Upon a type transition, \textit{Gemini}  identifies the block as highly unstable and  assigns high priority to it if its count is not "00". Otherwise, this block is assigned to low priority. Note that a leading block always keeps its counter's state. This policy eliminates the short, isolated type transitions.

\section{experimental methodology}
\label{sec_5_methodology}

We evaluate the performance of \textit{Gemini} using Gem5 \cite{Binkert2011}, integrated with detailed models of 3D-stacked DRAM and off-chip memory \cite{poremba2012}. The architectural parameters are summarized in Table~\ref{tbl_architecture_params}. The processor has 8 out-of-order cores with 16MB shared L2 cache. The DRAM cache is 1GB. We assume   both DRAM cache and the off-chip memory have the same latency,  but the DRAM cache has higher bandwidth by 8 times ($2\times$bus width, $2\times$channels, $2\times$banks). 

\subsection{Cache Organizations}
\label{sec_5_1_cache_organizations}

We compare \textit{Gemini} with the following designs. The key system configurations are listed in Table~\ref{tbl_sys_config}.

\begin{table} [b]
	\caption{\normalsize{\textbf{Architectural parameters}}}
	{\centering \ra{1.1}
		\begin{tabular}{@{}L{1.8cm}L{6.1cm}@{}}
			\toprule
			{\ }Processor & Out-of-order, 3.2GHz, 8 cores \\
			\hline
			{\ }L1 I/D Cache & 32KB I-cache, 32KB D-cache, \\
			& private, 4-way, 2-cycles\\
			\hline
			{\ }L2 Cache & 16MB, Shared, 8-way, 20-cycles, non-inclusive\\
			\hline
			{\ }Tag Cache & 32K entries, 8-way, 9-cycles\\
			\hline
			{\ }DRAM Cache & 1GB, 1.6GHz (DDR 3.2GHz), non-inclusive\\
			& 4 channels, 128 bits per channel, \\
			& 16 banks per rank, 2KB row buffer \\
			& tCAS-tRCD-tRP-tRAS 36-36-36-144 CPU cycles \\
			\hline
			{\ }\small{Main Memory} & 16GB, 800MHz (DDR 1.6GHz)\\
			& 2 channels, 64 bits per channel, \\
			& 8 banks per rank, 2KB row buffer \\
			& tCAS-tRCD-tRP-tRAS 36-36-36-144 CPU cycles \\
			\bottomrule
		\end{tabular}
	}
	\setlength{\abovecaptionskip}{3pt} 
	\setlength{\belowcaptionskip}{-6pt}
	\label{tbl_architecture_params}
\end{table}

\begin{table}
	\caption{\normalsize{\textbf{System configurations}}}
	{\centering \ra{1.1}
		\begin{tabular}{@{}l |l l l@{}}
			\toprule
			\multicolumn{1}{c}{} & BEAR & L-H & Gemini  \\
			\hline
			Mapping Policy & static & dynamic & static+dynamic \\
			Associativity & 1 & 14 & 16 \\
			Replacement & BAB & CLOCK & RV-CLOCK\\
			Tag Prefetch & neighboring tag & tag batch & tag batch\\
			Write Probe & DCP & DCP & DCP\\
			\hline
		\end{tabular}
	}
	\setlength{\abovecaptionskip}{2pt} 
	\setlength{\belowcaptionskip}{-8pt}
	\label{tbl_sys_config}
\end{table}

\textbf{Loh-Hill Cache} \cite{loh11,loh12}  We enhance the  Loh-Hill cache with tag cache in SRAM as a baseline, which is refereed to as enchanced-LH cache.  The associativity  is reduced from the original 29 to 14 with the considerations of tag batch size, for the lower tag access latency. In our implementation, the tags and the data of the same set are placed in the same DRAM row. The tags are batched in a per-set manner, and are read together in a single request. 

\textbf{BEAR cache} \cite{chou15}  As direct-mapped DRAM cache,  BEAR is compared with our design. It uses a Bandwidth Aware Bypass (BAB) scheme to improve bandwidth efficiency upon miss fill. The tags and the data are placed in DRAM rows in an interleaving manner. A read request fetches the data, as well as the corresponding tag and the next neighboring tag with one additional burst.  To be fair,  the SRAM tag caches are set to be same size for the enhanced LH cache, BEAR and Gemini. In addition, the DRAM Cache Presence (DCP) bit \cite{chou15}  is included in these 3 DRAM cache designs to reduce the bandwidth consumption. 


\subsection{Workloads}
\label{sec_5_2_workloads}

\begin{table}
	\caption{\normalsize{\textbf{Workloads}}}
	{\centering \ra{1.1}
		\begin{tabular}{@{}c@{} |@{}l@{} l@{} l@{}}
			\toprule
			\multicolumn{4}{c}{Rate Workloads}  \\
			\hline
			CD & {\ }R1: wupwise$\times$8 & {\ }{\ }R2: lucas$\times$8 & {\ }{\ }R3: gap$\times$8  \\
			& {\ }R4: apsi$\times$8 & {\ }{\ }R5: cactusADM$\times$8 &  {\ }{\ }R6: lbm$\times$8 \\
			\hline
			LD & {\ }R7: vpr$\times$8 & {\ }{\ }R8: bzip2$\times$8 & {\ }{\ }R9: soplex$\times$8 \\
			& {\ }R10: omnetpp$\times$8 & {\ }{\ }R11: astar$\times$8 &  {\ }{\ }R12: xalan$\times$8 \\
			\hline
			BF & {\ }R13: equake$\times$8 & {\ }{\ }R14: mgrid$\times$8 & {\ }{\ }R15: gcc$\times$8 \\
			& {\ }R16: libquantum$\times$8 & & \\
			\hline
			NB & {\ }R17: GemsFDTD$\times$8 & {\ }{\ }R18: milc$\times$8 & {\ }{\ }  \\
			\hline
			\hline
			\multicolumn{4}{c}{Mixed Workloads}  \\
			\hline
			CD & \multicolumn{3}{@{}l}{{\ }M1:wupwise,luca,gap,apsi,cactusADM,lbm,wupwise,lbm}\\
			\hline
			LD & \multicolumn{3}{@{}l}{{\ }M2:vpr,bzip2,soplex,omnetpp,astar,xalan,vpr,xalan}\\
			\hline
			BF & \multicolumn{3}{@{}l}{{\ }M3:{\{}equake,mgrid,gcc,libquantum{\}}$\times$2}\\
			\hline
			NB & \multicolumn{3}{@{}l}{{\ }M4:{\{}milc,GemsFDTD{\}}$\times$4}\\
			\hline
			CD+LD & \multicolumn{3}{@{}l}{{\ }M5:\small{wupwise,gap,apsi,cactusADM,vpr,bzip2,soplex,omnetpp}}  \\
			\hline
			CD+BF & \multicolumn{3}{@{}l}{{\ }M6:\small{gap,apsi,cactusADM,lbm,equake,mgrid,gcc,libquantum}}  \\
			\hline
			CD+NB & \multicolumn{3}{@{}l}{{\ }M7:\small{lucas,apsi,cactusADM,lbm,{\{}milc,GemsFDTD{\}}$\times$2}}  \\
			\hline
			LD+BF & \multicolumn{3}{@{}l}{{\ }M8:\small{soplex,omnetpp,astar,xalan,equake,mgrid,gcc,libquantum}}  \\
			\hline
			LD+NB & \multicolumn{3}{@{}l}{{\ }M9:\small{vpr,bzip2,astar,xalan,{\{}milc,GemsFDTD{\}}$\times$2}}  \\
			\hline
			BF+NB & \multicolumn{3}{@{}l}{{\ }M10:\small{equake,mgrid,gcc,libquantum,{\{}milc,GemsFDTD{\}}$\times$2}}  \\
			\hline
		\end{tabular}
	}
	\setlength{\abovecaptionskip}{3pt} 
	\setlength{\belowcaptionskip}{-15pt}
	\label{tbl_workloads}
\end{table}

We evaluate the three designs with 18 memory intensive benchmarks from the SPEC CPU2000 and the SPEC CPU2006 benchmark suites. The benchmarks are classified based on two metrics: spatial locality and cache contention. The spatial locality is measured as $N_{leading}/N_{total}$, where $N_{leading}$ and $N_{total}$ are the amount of leading blocks' accesses  and the amount of  the total accesses respectively. The cache contention is evaluated as the DRAM cache's hit rate\footnote{The hit rate is taken from the rate mode workloads running on the direct-mapped cache.}. 

\begin{figure*}
	\centering
	\includegraphics[width=0.99\textwidth]{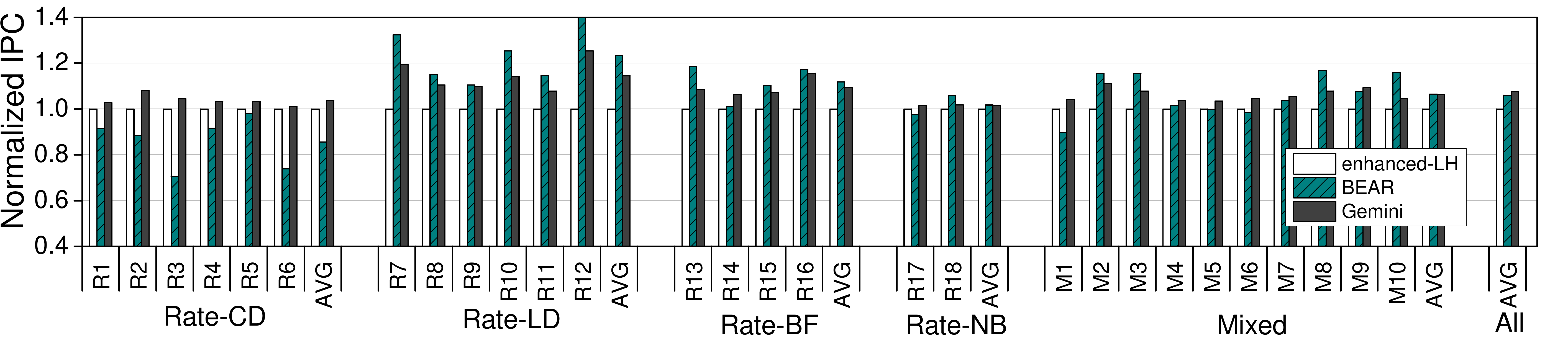}
	\setlength{\abovecaptionskip}{0pt} 
	\setlength{\belowcaptionskip}{0pt}
	\caption{\normalsize{\textbf{Normalized IPC. The results are normalized based on enhanced-LH's IPC.}}}
	\label{fig_ipc}
\end{figure*}

These benchmarks can be classified into four categories: contention-dominated (CD), locality-dominated (LD), both-friendly (BF), and non-beneficial (NB). The CD benchmarks have high cache demands and good spatial locality. The set-associative cache is friendly to these benchmarks, as it can improve the cache space utilization, and benefits from the tag cache due to the spatial locality. In contrast, the LD benchmarks have high hit rate, but poor spatial locality. The BEAR works well for these benchmarks because the hit latency has been optimized. The BF benchmarks do not have much cache demand with good spatial locality. Both basic cache organizations can perform well on these benchmarks. The NB benchmarks have poor spacial locality and high miss rate, and no design can work well for them.

The Table~\ref{tbl_workloads} shows 28 workloads in our experiments. There are 18 workloads in rate mode (four categories, called Rate-CD, Rate-LD, Rate-BF, Rate-NB in the following results analysis), where all the cores execute the same benchmark. We also evaluate 10 mixed workloads. The intra-category and inter-category combinations are both included. For each workload, we the simulate 1 billion instructions on each core after fast-forwarding the first 10 billion instructions 


\section{Results}
\label{sec_6_results}


\subsection{Performance}
\label{sec_6_1_performance}

We normalize the IPC (instructions per cycle) of three designs to that of enhanced-LH Cache, and show the results in Figure~\ref{fig_ipc}. For Rate-CD workloads, \textit{Gemini} outperforms the direct-mapped cache, BEAR,  17.5\% on average and up to 33\% . The speedup of \textit{Gemini} mainly comes from hit rate improvement due to its dynamic mapping of following blocks. The Rate-LD workloads have poor spatial locality, which leads to high miss rate on tag cache. For these workloads, enhanced-LH Cache performs worst, as it serializes tag and data accesses. By applying static mapping to leading blocks, \textit{Gemini} outperforms enhanced-LH Cache by 18\% on average and up to 23\%. The Rate-BF workloads have small working sets and high spatial locality, which narrows the performance gap between different designs. On average, \textit{Gemini} achieves similar performance comparing with BEAR (1 vs. 1.02), and outperforms enhanced-LH Cache by 7\%. The Rate-NB workloads offer low temporal locality and large working sets. None of the designs perform well on these workloads. For the Mixed workloads, Gemini consistently outperforms enhanced-LH Cache. In four out of ten combinations (\textit{M2, M3, M8, M10}), Gemini achieves lower IPC comparing to BEAR, with less than 6\% performance difference.

\subsection{Requests Breakdown}
\label{sec_6_2_request_breakdown}

\begin{figure*}
	\captionsetup[subfigure]{labelformat=empty}
	\centering
	\begin{subfigure}[b]{0.99\textwidth}
		\includegraphics[width=1\linewidth, height=1.8in]{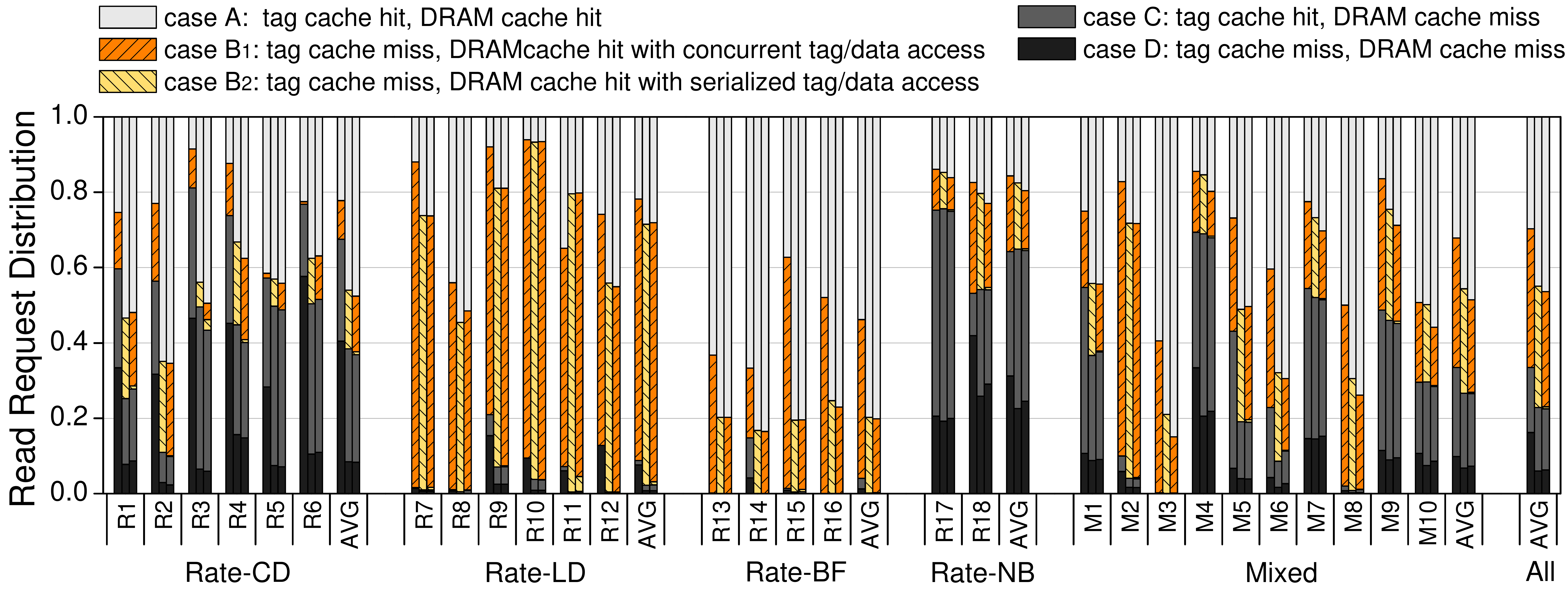}
		\setlength{\abovecaptionskip}{0pt} 
		\setlength{\belowcaptionskip}{-5pt}
		\caption{\normalsize{\textbf{Figure 10: Read request distribution based on data access path. From left to right, the three bars in each group are the results of BEAR Cache, enhanced-LH Cache, and \textit{Gemini} Cache respectively.}}}
		\label{fig_read_request_distribution_sub} 
	\end{subfigure}
	\vspace{8pt}
	
	\begin{subfigure}[b]{0.99\textwidth}
		\includegraphics[width=1\linewidth, height=1.2in]{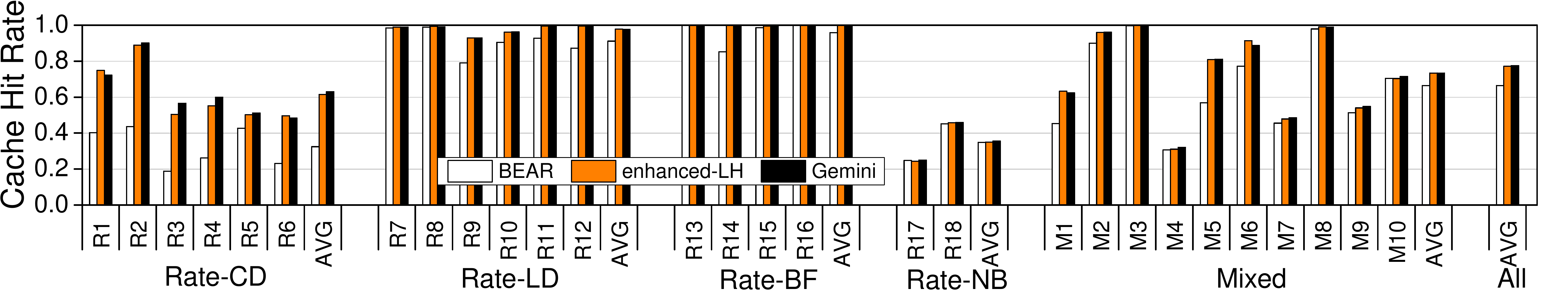}
		\setlength{\abovecaptionskip}{-5pt} 
		\setlength{\belowcaptionskip}{0pt}
		\caption{\normalsize{\textbf{Figure 11: DRAM cache hit rate}}}
		\label{fig_hit_rate}
	\end{subfigure}
	\label{fig_read_request_distribution} 
\end{figure*}
\setcounter{figure}{11}

%

To investigate how data mapping policies affect cache performance, we study the read request distributions\footnote{In this study, we mainly focus on read requests, as the write requests are not in the critical path.} based on data access path. As we have discussed in section~\ref{sec_2_2_latency}, data requests can go through different paths in the cache system. Case A: the request hits the tag cache, then hits the DRAM cache; Case B$_1$: the request misses the tag cache, but hits the DRAM cache with concurrent tag and data accesses; Case B$_2$: the request misses the tag cache, but hits the DRAM cache with serialized tag and data accesses; Case C: the tag cache is hit, and the request heads directly to the main memory to fetch the data; Case D: the tag cache is missed, the request first checks the tag in DRAM cache, then fetches data from main memory. 

The distributions of different cases are shown in Figure 10. The tag cache hit rate can be calculated by adding the portions of cases A and C, and the DRAM cache hit rate equals to the sum of cases A, B$_1$, and B$_2$. For clarity, we also show the DRAM cache hit rate in Figure 11.

For Rate-CD workloads, BEAR achieves relatively low hit rate because its static data mapping limits replacement flexibility. Comparing with BEAR, \textit{Gemini} doubles the average hit rate, from 32\% to 63\%. It is important to note that, \textit{Gemini} achieves the similar hit rate as enhanced-LH Cache does (the hit rate difference is less than 5\%). This implies, applying dynamic mapping for following blocks can offer good replacement flexibility, and the leading blocks can be mapped statically without introducing hit rate reduction. In addition, \textit{Gemini} successfully removes the majority of serialized tag and data accesses (case B$_2$), which brings 3.5\% speedup comparing with enhanced-LH Cache (see figure~\ref{fig_ipc}).

The Rate-LD workloads have small working sets that can be accommodated into the DRAM cache. As a result, all the three designs achieve a high hit rate. However, these workloads offer poor spatial locality, which leads to an average of 72\% miss rate in the tag cache, and introduces a large ratio of leading blocks. The cache system's performance is dominated by the DRAM cache's hit latency. As we learn from Figure 10, 70\% of the cache hits involve tag fetching. For enhanced-LH Cache, all these requests need to be done in a serialized data-after-tag fashion. In comparison, \textit{Gemini} localizes leading blocks at their statically mapped positions. On average, 69\% cache hits are served by concurrent tag and data fetching, which is very close to BEAR.

The \textit{Rate-BF} workloads' working set can also fit into the DRAM cache, while offering high spatial locality. The tag batching mechanism helps \textit{Gemini} and enhanced-LH Cache obtain 80\% hit rate in tag cache. In contrast, the neighboring tag prefetching technique used by BEAR achieves 57\% hit rate in tag cache. But the difference in tag cache's hit rate cannot bring more benefits for the set-associative cache (enhanced-LH Cache), because the direct-mapped cache (BEAR) offers same latency as long as the DRAM cache is hit. On the other hand, there is still a small portion of requests need to access the tags in DRAM cache. These requests slightly degrade the performance of enhanced-LH Cache by about 8\%.

The \textit{Rate-NB} workloads have large working set and the poor temporal locality. This leads to a low hit rate in DRAM cache, which limits the influence of data mapping policies, thus all the three designs have the similar request distributions. In Figure~\ref{fig_ipc}, we observe that all the designs get the similar IPC under these workloads. 


The \textit{Mixed} workloads \textit{M1, M5, M6} offer optimization space in both hit rate improvement and hit latency reduction. For these workloads, Gemini improves the average hit rate by 28\% comparing with BEAR, and eliminates over 95\% of the serialized tag and data accesses. It outperforms BEAR and enhanced-LH Cache by 8\% and 5\% on average. The workloads \textit{M2, M3, M8} have high hit rate in DRAM cache (over 90\%). BEAR is the best performed design for these workloads, followed by Gemini. Due to their concurrent tag and data accesses, these two designs achieve 15\% and 9\% higher IPC compared with enhanced-LH Cache.

\subsection{The Impact of Tag Fetching Mode on Hit Latency}
\label{sec_6_3_cache_hit_latency}

For \textit{LD} workloads (\textit{R7}-\textit{R12}, \textit{M2}) with poor spatial locality, the cache system's performance is sensitive to the access latency of leading blocks. The enhanced-LH Cache is unaware of leading or following blocks and applies unified dynamic data mapping. In comparison, \textit{Gemini} apply static mapping for the leading blocks. Note that both \textit{Gemini} and BEAR can concurrently fetch tag and data, but the underlying mechanism is different. \textit{Gemini} accesses a leading block with two parallel DRAM cache requests: one for the tag batch, and the other for the data. BEAR co-locates each tag data pair, thus the whole tag-and-data entity can be fetched in a single request (with one additional burst). 

\begin{figure}
	\centering
	\includegraphics[width=0.40\textwidth, height=1.2in]{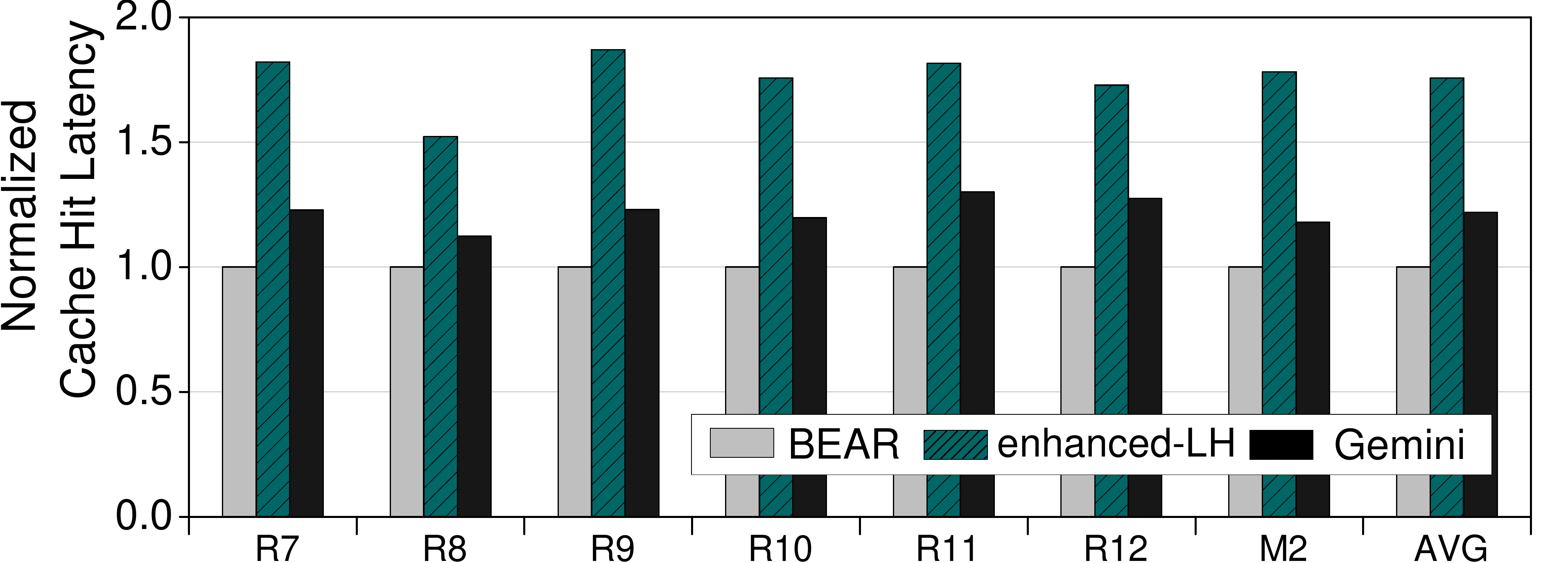}
	\setlength{\abovecaptionskip}{0pt} 
	\setlength{\belowcaptionskip}{-15pt}
	\caption{\normalsize{\textbf{Hit latency of Locality-Dominated workloads. The results are normalized based on BEAR's hit latency.}}}
	\label{fig_ld_hit_latency}
\end{figure}

To illustrate the impact of different tag fetching modes, we show their hit latency in Figure~\ref{fig_ld_hit_latency}. BEAR has the lowest hit latencies. In comparison, enhanced-LH Cache's hit latency is 1.75X on average and 1.87X to the maximum. Due to the queuing delay of \textit{Gemini}, not all the tag and data requests can be done at exactly the same time. As a result, \textit{Gemini}'s hit latency is 1.22X comparing to BEAR. For the \textit{LD} workloads, the leading blocks' access latency directly affects overall system performance. In Figure~\ref{fig_ipc}, we can see that BEAR achieves 23\% higher IPC (on average) compared with enhanced-LH Cache. \textit{Gemini} reduces the performance gap down to 6\% with its sophisticated data mapping policy.

\subsection{The Impact of Hit Rate Improvement on Main Memory Queuing Delay}
\label{sec_6_4_queuing_delay}

High hit rate benefits system performance in two-fold. Firstly, the requests that hit the cache can enjoy the high bandwidth of the die-stacking DRAM.  Secondly, high hit rate can help to reduce the main memory's access latency. Due to the filtering effect of DRAM cache, a large portion of off-chip accesses are eliminated. The reduced access frequency leads to a relatively low queuing delay on main memory. 

\begin{figure}
	\centering
	\includegraphics[width=0.45\textwidth]{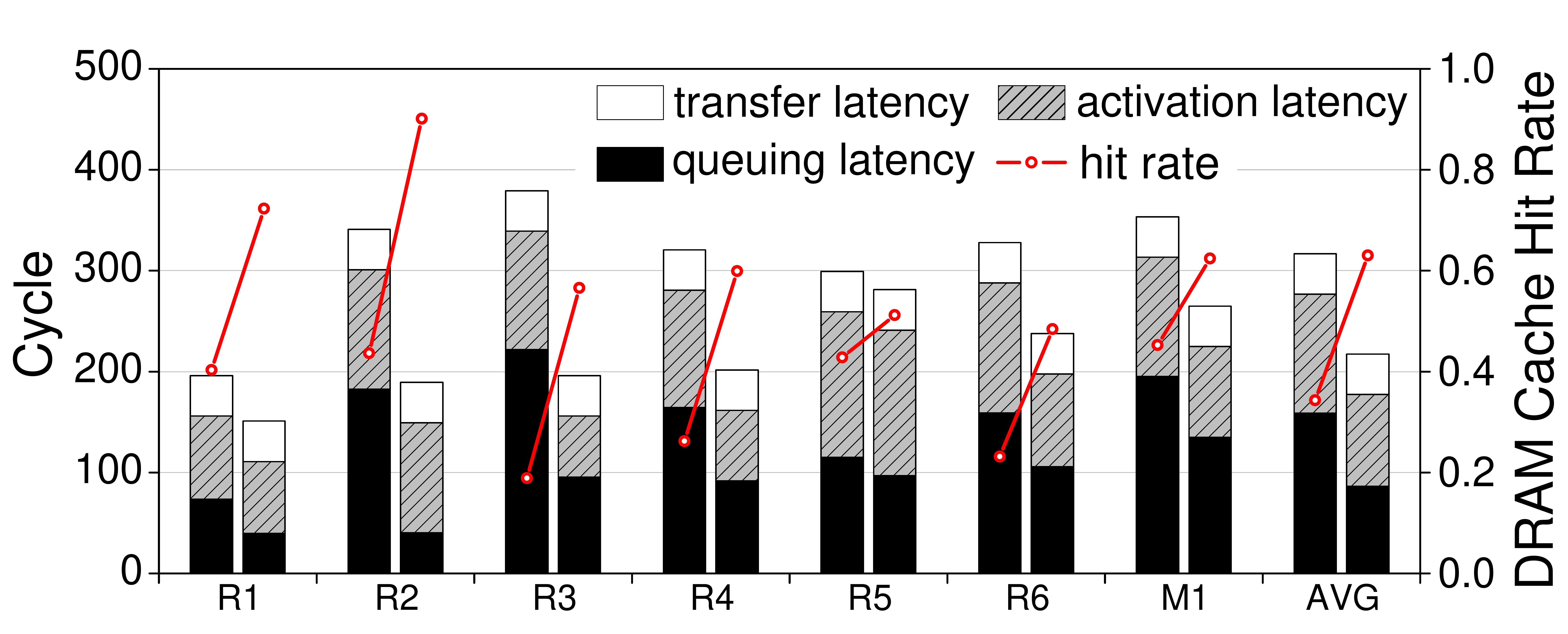}
	\setlength{\abovecaptionskip}{0pt} 
	\setlength{\belowcaptionskip}{-15pt}
	\caption{\normalsize{\textbf{The impact of DRAM cache hit rate on main memory queuing delay. From left to right, the two bars in each group are the results of BEAR and \textit{Gemini}. }}}
	\label{fig_dc_hit_rate_vs_memory_queuing_delay}
\end{figure}

Figure~\ref{fig_dc_hit_rate_vs_memory_queuing_delay} shows main memory queuing delay of the CD workloads. We compare the results of BEAR against \textit{Gemini} to study the impact of hit rate. From left to right, the two bars in each group are the results of BEAR Cache and \textit{Gemini} Cache respectively. We also show the hit rate with red lines. As the figure shows, the low hit rate of DRAM cache makes the CD workloads suffer a severely high queuing delay. In the worst case, i.e., \textit{R3}, the queuing delay contributes up to 58\% of the overall memory access latency. By improving the DRAM cache's hit rate, \textit{Gemini} reduces the queuing delay by 49\% on average and 78\% at the maximum.

\subsection{High Frequency Type Variation Filter}
\label{sec_6_6_efficiency_of_filter}

\begin{figure}
	\centering
	\includegraphics[width=0.45\textwidth]{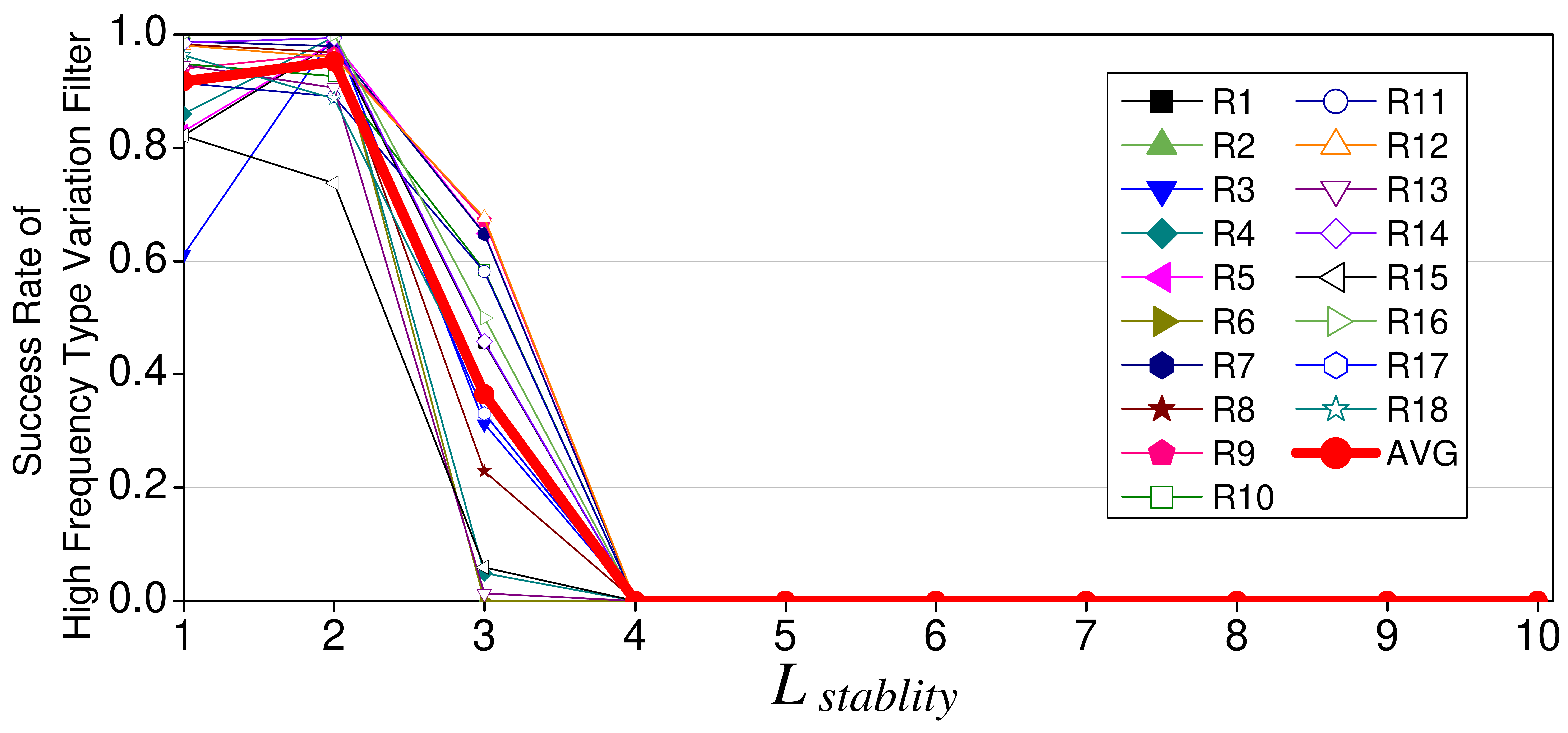}
	\setlength{\abovecaptionskip}{0pt} 
	\setlength{\belowcaptionskip}{-15pt}
	\caption{\normalsize{\textbf{The success rate of the High Frequency Type Variation Filter}}}
	\label{fig_filter_success_rate}
\end{figure}

Gemini relies on the \textit{high frequency type variation filter} to track the unstable data blocks. Figure~\ref{fig_filter_success_rate} shows the percentage of the identified cases with different \textit{L}$_{stable}$. As we can see, the filter has high resistance against short \textit{L}$_{stable}$. The filtering rate of \textit{L}$_{stable}$=1, 2 are 92\% and 95\% respectively. Meanwhile, it pass through all the segments with \textit{L}$_{stable}>3$.

\section{Related Work}
\label{sec_7_related_work}



\noindent
\textbf{Set-associative Cache}. Loh-Hill \textit{et al.} \cite{loh11,loh12} propose to organize each DRAM row (2 KB) as a 29-way cache set, which provides good replacement flexibility. 
The tag and data are stored in the same row, and the data is requested right after the tag read to improve row buffer efficiency. Unison Cache \cite{jevdjic2014} and Bi-Modal Cache \cite{gulur14} increase the cache line granularity to exploit the spatial locality, and they rely on the set-associative structure to mitigate the cache conflicts.

\noindent
\textbf{Direct-mapped Cache}. Alloy Cache \cite{qureshi12} breaks the data-after-tag serialization by statically mapping each data block to the DRAM cache. It also combines each tag with its corresponding data into a single entity to avoid the standalone tag read request, which introduces one additional burst. CAMEO \cite{chou2014} improves this design by eliminating the data copy between the direct-mapped cache and the main memory. The in-memory data that mapped to the same cache line, as well as the data that currently occupies the line, are identical to each other, and contend for the cache line via swapping, instead of inserting the data copy. 

These studies present different optimizations on the set-associative DRAM cache or the direct-mapped DRAM cache. In this paper, we reveal that different data blocks can significantly impact on cache hit latency, and then propose a partial direct-mapped cache design to gain both set-associative structure's high hit rate, and the direct-mapped structure's low hit latency. 

\section{Conclusion}
\label{sec_8_conclusion}

In this paper, we have presented \textit{Gemini}, a partial direct-mapped die-stacked DRAM cache. \textit{Gemini} cache classifies data into leading blocks and following blocks and places them with static mapping and dynamic mapping respectively in a unified set-associative structure. The key idea is inspired by the observations that different blocks have different impact on the cache hit rate and cache hit latency. Experimental results demonstrate that \textit{Gemini} cache can narrow the hit latency gap with direct-mapped cache significantly, from 1.75X to 1.22X on average, and can achieve comparable hit rate with set-associative cache. Compared with the state-of-the-art baselines, i.e., BEAR cache and Loh-Hill cache, \textit{Gemini} improves the IPC by up to 33\% and 20\% respectively.

\bibliographystyle{ACM-Reference-Format}
\bibliography{main}

\end{document}